\documentclass[12pt]{article}

\usepackage{url}
\usepackage{amsmath,amsthm}

\usepackage{amsfonts}
\usepackage{natbib}

\usepackage[dvips]{graphicx}

\makeatletter
\newcommand{\@makemycaption}[2]{%
\vspace{10pt}%
{\textbf{#1}:#2\par}%
}
\renewcommand{\figure}{%
\let\@makecaption\@makemycaption\@float{figure}}
\renewcommand{\table}{%
\let\@makecaption\@makemycaption\@float{table}}
\makeatother

\date{}

\newcommand{\eps}{\varepsilon}
\newcommand{\To}{\rightarrow}

\numberwithin{equation}{section}

\begin{document}

\begin{center}
  \textbf{\large Efficient adaptive designs with mid-course sample size
  adjustment in clinical trials}

\bigskip

\textbf{Jay Bartroff}\\ Department of Mathematics, University of Southern California, Los
 Angeles, CA, USA {\renewcommand{\thefootnote}{}\footnote{Address
 correspondence to  Jay Bartroff, Department of Mathematics, USC, 3620 S Vermont Ave KAP 108, Los Angeles, CA, USA; E-mail: bartroff@usc.edu}}

\medskip

\textbf{Tze Leung Lai}

Department of Statistics, Stanford University, Stanford, CA, USA
\end{center}

\bigskip

\noindent\textbf{\small Abstract:} {\small Adaptive designs have been proposed for clinical
  trials in which the nuisance parameters or alternative of interest
  are unknown or likely to be misspecified before the trial.  Whereas
  most previous works on adaptive designs and mid-course sample size
  re-estimation have focused on two-stage or group sequential designs
  in the normal case, we consider here a new approach that involves at
  most three stages and is developed in the general framework of
  multiparameter exponential families. Not only does this approach
  maintain the prescribed type I error probability, but it also
  provides a simple but asymptotically efficient sequential test whose
  finite-sample performance, measured in terms of the expected sample
  size and power functions, is shown to be comparable to the optimal
  sequential design, determined by dynamic programming, in the
  simplified normal mean case with known variance and prespecified
  alternative, and superior to the existing two-stage designs and also
  to adaptive
  group sequential designs when the alternative or nuisance parameters
  are unknown or misspecified.}

\vspace{1.5in}



\newpage

\section{INTRODUCTION}
\label{sec:intro}
In standard clinical trial designs, the sample size is determined by
the power at a given alternative (e.g., treatment effect). In
practice, especially for new treatments about which there is little
information on the magnitude and sampling variability of the treatment
effect, it is often difficult for investigators to specify a realistic
alternative at which sample size determination can be based.
Therefore, the problem of sample size re-estimation based on an
observed treatment difference at some time before the prescheduled end
of the trial has attracted considerable attention during the past
decade; see e.g.\ Jennison and Turnbull~\citeyearpar[Section
~14.2]{Jennison00}, Shih~\citeyearpar{Shih01} and Whitehead et
al.~\citeyearpar{Whitehead01}.  Moreover, there are concerns from the
regulatory perspective regarding possible inflation of the type I
error probability when such sample size adjustments are used in
pharmaceutical trials. For normally distributed outcome variables,
Proschan and Hunsberger~\citeyearpar{Proschan95}, Fisher~\citeyearpar{Fisher98},
Posch and Bauer~\citeyearpar{Posch99}, and Shen and Fisher~\citeyearpar{Shen99} have
proposed ways to adjust the test statistics after mid-course sample
size modification so that the type I error probability is maintained
at the prescribed level. Jennison and Turnbull~\citeyearpar{Jennison03} gave
a general form of these methods and showed that they performed
considerably worse than group sequential tests. Tsiatis and
Mehta~\citeyearpar{Tsiatis03} independently came to the same conclusion,
pointing out their inefficiency because the adjusted test statistics
are not sufficient statistics. It is possible to adhere to efficient
generalized likelihood ratio statistics in a mid-course adaptive
design if one uses the non-normal sampling distribution (due to the
mid-course adaptation) of the test statistic, instead of ignoring the
nonnormality and thereby resulting in type I error inflation. A way to
do this was proposed by Li et al.~\citeyearpar{Li02}, but it was shown by
Turnbull~\citeyearpar{Turnbull06} to be relatively inefficient compared to
group sequential tests. Jennison and Turnbull~\citeyearpar{Jennison06a}
recently introduced adaptive group sequential tests that choose the
$j$th group size and stopping boundary on the basis of the cumulative
sample size $n_{j-1}$ and the sample sum $S_{n_{j-1}}$ over the first
$j-1$ groups, and that are optimal in the sense of minimizing a
weighted average of the expected sample sizes over a collection of
parameter values subject to prescribed error probabilities at the null
and a given alternative hypothesis. They also showed how the
corresponding optimization problem can be solved numerically by using
the backward induction algorithms for ``optimal sequentially planned''
designs developed by Schmitz~\citeyearpar{Schmitz93}. Jennison and
Turnbull~\citeyearpar{Jennison06b} found that standard (non-adaptive) group
sequential tests with the first stage chosen optimally are nearly as
efficient as their optimal adaptive counterparts that are considerably
more complicated, and we use these as a benchmark for our comparisons
in Section~\ref{sec:sim}.

With the goal of achieving similar efficiency in more complicated
situations where the alternative of interest and/or nuisance
parameters are not known, we give in Section~\ref{sec:1param} a simple
adaptive test which updates the sample size after the initial stage
by using estimates of the unknown parameters and adjustments for
the uncertainty of these estimates. This is done first for the
one-parameter case in Section~\ref{subsec:1param} and extended to the
multiparameter setting in Section~\ref{sec:multi}.  These tests
usually terminate at the first or second stage, but allow the
possibility of a third stage to account for uncertainties in the
second-stage sample size estimate.  The tests control the type I error
probability and have power close to the uniformly most powerful fixed
sample test.  Section~\ref{sec:sim} gives a comprehensive simulation
study, which is the first of its kind, of the adaptive tests in the
aforementioned references and compares them with the adaptive tests
developed in Section~\ref{sec:1param} and with fixed sample size and
standard group sequential tests having the same minimum and maximum
sample sizes. A thorough evaluation of the performance of these tests
is presented, involving the power, mean number of stages, and the
mean, 25th, 50th, and 75th percentiles of the sample size distribution
under a wide range of alternatives, subject to the prescribed
constraints on type I error probability and first-stage and maximum
sample sizes. Section~\ref{subsec:varknown} also compares the proposed
adaptive test with the benchmark optimal adaptive test of Jennison and
Turnbull~\citeyearpar{Jennison06a,Jennison06b}, and the variance unknown case
is considered in Section~\ref{subsec:varunknown}. An example from the
National Heart, Lung and Blood Institute Coronary Intervention Study
is given in Section~\ref{subsec:2bino}.  Section~\ref{sec:disc} gives
some concluding remarks.

\section{EFFICIENT ADAPTIVE TESTS WITH THREE OR FEWER STAGES}
\label{sec:1param}

In this section we consider one-sided tests of the null hypothesis
$H_0 : \theta \leq \theta_0$ on the natural parameter $\theta$ in a
one-parameter exponential family $f_\theta (x) = e^{\theta x - \psi
  (\theta)}$ of densities with respect to some measure on the real
line. Let $X_1, X_2, \ldots$ denote the successive observations, and
let $S_n = X_1 + \ldots + X_n$. A sufficient statistic based on $(X_1,
\ldots, X_n)$ is $\bar X_n = S_n / n$, and the
maximum likelihood estimate of $\theta$ is
$\widehat{\theta}_n=(\psi')^{-1}(\bar X_n)$. The special case of
normal $X_i$ with mean $\theta$ and known variance $1$ is widely used
in the literature on sample size re-estimation as a prototype which
can be used to
approximate more complicated situations via the central limit
theorem, as in the references in Section~\ref{sec:intro}.

In practice, there is an upper bound $M$ on the allowable sample size
for a clinical trial because of funding and duration constraints and
because there are other trials that compete for patients,
investigators and resources. The re-estimated sample size in two-stage
designs has to be restricted within this bound; see, e.g., Li et
al.~\citeyearpar[p.~283]{Li02}. Lai and Shih~\citeyearpar[p.~511]{Lai04} have
pointed out that $M$ implies constraints on the alternatives that can
be considered in power calculations to determine the sample size.
Specifically, by the Neyman-Pearson lemma, the fixed sample size (FSS)
test that rejects $H_0$ if $S_M \geq c_{\alpha, M}$ has maximal power
at any alternative $\theta > \theta_0$, and in particular at the
alternative $\theta_1$ at which the FSS test has prescribed power $1 -
\widetilde \alpha$. Here $c_{\alpha, n}$
denotes the critical value of the level-$\alpha$ FSS test based on a
sample of size $n$, i.e., $\mbox{pr}_{\theta_0} \{S_n \geq c_{\alpha,
  n}\} = \alpha$. Typical sample size re-estimation procedures in the
literature (see, e.g., the references in Section~\ref{sec:intro})
first use the initial sample of size $m$, which is some fraction of
$M$, to provide an estimate $\widehat \theta_m$ of $\theta$ and then
evaluate the sample size of the FSS test that has conditional power $1
- \widetilde \alpha$ given the alternative $\widehat \theta_m$,
assuming that $\widehat \theta_m >\theta_0$. This results in a
two-stage procedure, which does not incorporate the sampling
variability of the estimate $\widehat \theta_m$. A simple way to make
``uncertainty adjustments'' in the above procedure that attempts to
``self-tune'' itself to the actual $\theta$ value is to allow the
possibility of not stopping at the second stage when $H_0$ is not
rejected, by including a third (and final) stage with total sample
size $M$.

\subsection{An Efficient Test of $H_0$ with At Most Three Stages}
\label{subsec:1param}

To test $H_0 : \theta \leq \theta_0$ at significance level $\alpha$,
suppose no fewer than $m$ but no more than $M$ observations are to be
taken.  Let $\theta_1$ be the alternative ``implied'' by $M$, in the
sense that $M$ can be determined as the sample size of the
level-$\alpha$ Neyman-Pearson test with power $1 - \widetilde \alpha$
at $\theta_1$. Alternatively, $\theta_1$ can be specified separately
from $M$ as a clinically relevant or realistic anticipated effect size
based on prior experimental, observational, or theoretical evidence,
if such information is available. A fundamental result in sequential
testing theory is that Wald's sequential probability ratio test (SPRT)
of the simple hypotheses $\theta=\theta'$ vs.\ $\theta=\theta''$ has
the smallest expected sample size at $\theta=\theta'$ and $\theta''$
among all tests with the same or smaller type~I and II error
probabilities; see Reference~\citeyearpar{Chernoff72}. Moreover, letting
$\alpha$ and $\widetilde{\alpha}$ denote the the type I and II error
probabilities and $T(\theta',\theta'')$ be the sample size of the
SPRT, Chernoff~\citeyearpar[p.~66]{Chernoff72} has derived the approximations
\begin{equation}E_{\theta''}(T(\theta',\theta''))\approx
  |\log\alpha|/I(\theta'',\theta'),\quad E_{\theta'}(T(\theta',\theta''))\approx
  |\log\widetilde{\alpha}|/I(\theta',\theta''),\label{eq1}
\end{equation} where
$$I(\theta, \lambda) = E_\theta [\log \{f_\theta (X_i) / f_\lambda (X_i)\}]
= (\theta - \lambda) \psi^\prime (\theta) - \{\psi(\theta) - \psi
(\lambda)\}
$$
is the Kullback-Leibler information number. To test the one-sided
hypothesis $H_0:\theta\le\theta_0$, suppose that we use the maximum
likelihood estimator $\widehat{\theta}_m$ from the first stage of the
study in place of the alternative $\theta''$ in (\ref{eq1}) with $\theta'=\theta_0$,
in the event $\widehat{\theta}_m>\theta_0$. Then the first relation in
(\ref{eq1}) suggests that an efficient second-stage sample size would
be around $|\log\alpha|/I(\widehat{\theta}_m,\theta_0)$. On the other
hand, if $\widehat{\theta}_m\le\theta_0$, then we can consider the
possibility of stopping due to futility by choosing
$\theta'=\widehat{\theta}_m$ and $\theta''=\theta_1$ in the SPRT, so
the second relation in (\ref{eq1}) suggests
$|\log\widetilde{\alpha}|/I(\widehat{\theta}_m,\theta_1)$ as an
efficient second-stage sample size. Adjusting for the sampling
variability in $\widehat{\theta}_m$ by inflating by the factor $1+\rho_m$, we therefore define the second
stage sample size
\begin{equation} n_2 =m\vee\{M\wedge\lceil (1 + \rho_m) n(\widehat
  \theta_m)\rceil\},\label{eq2}\end{equation} where $\rho_m>0$,
$\vee$ and $\wedge$ denote maximum and minimum, respectively, $\lceil
x \rceil$ denotes the smallest integer $\geq x$ (and $\lfloor x
\rfloor$ denotes the largest integer $\leq x$), and
\begin{equation}n(\theta)=\frac{|\log\alpha|}{I(\theta,\theta_0)}\wedge
  \frac{|\log\widetilde{\alpha}|}{I(\theta,\theta_1)}, \label{eq3}
  \end{equation} which is an approximation to Hoeffding's~\citeyearpar{Hoeffding60} lower
  bound for the expected sample size $E_\theta (T)$ of a test that has
  type I error probability $\alpha$ at $\theta_0$ and type II error
  probability $\widetilde{\alpha}$ at $\theta_1$. Note that (\ref{eq2}) includes the cases $n_2=m$ and
  $n_2=M$ associated with using just one or two stages. Moreover, the
  stopping rule defined below by (\ref{eq4})-(\ref{eq6}) allows the
  possibility of stopping after the first or second stage. Therefore,
  the actual number of stages used by the ``three-stage'' test is in
  fact a random variable taking the values 1, 2, 3.    
  
The three-stage test uses rejection and futility boundaries similar to
those of the efficient group sequential tests introduced by Lai and
Shih~\citeyearpar{Lai04}.  Letting $n_i$ denote the total sample size at the
$i$th stage, the test stops at stage $i\le2$ and rejects $H_0$ if
\begin{equation} n_i<M,\quad \widehat \theta_{n_i} > \theta_0, \quad \mbox{and} \quad n_i
I(\widehat \theta_{n_i}, \theta_0) \geq b ,\label{eq4}\end{equation}
where $n_1 = m$ and $n_2$ is given by (\ref{eq2}). The test stops at stage $i \leq 2$ and accepts $H_0$ if
\begin{equation}
n_i<M,\quad\widehat \theta_{n_i} < \theta_1, \quad \mbox{and} \quad n_i
I(\widehat \theta_{n_i}, \theta_1) \geq \widetilde b . \label{eq5}
\end{equation} It rejects $H_0$ at stage $i=2$ or $3$ if
\begin{equation}
n_i=M,\quad \widehat \theta_M > \theta_0, \quad \mbox{and} \quad M
I(\widehat \theta_M, \theta_0) \geq c,\label{eq6}
\end{equation}  accepting $H_0$ otherwise. Letting $0 < \varepsilon, \widetilde \varepsilon <1$, define
the thresholds $b, \widetilde b$, and $c$ by the equations
\begin{eqnarray}
&&\mbox{pr}_{\theta_1} \{\mbox{(\ref{eq5}) occurs for} \ i=1 \
\mbox{or} \ 2\} =
\widetilde \varepsilon \widetilde \alpha, \label{eq7}\\
&&\mbox{pr}_{\theta_0} \{\mbox{(\ref{eq5})does not occur for} \
i\le 2, \mbox{ and (\ref{eq4}) occurs for} \ i = 1 \ \mbox{or} \ 2\} = \varepsilon \alpha ,\label{eq8}\\
&&\mbox{pr}_{\theta_0} \{\mbox{(\ref{eq4}) and (\ref{eq5}) do not occur
  for} \ i\le 2, \mbox{ and (\ref{eq6}) occurs}\} = (1-\varepsilon)\alpha. \label{eq9}
\end{eqnarray}  
Note that (\ref{eq8}) and (\ref{eq9}) imply that the type I error
probability is exactly $\alpha$, and we have found in our simulations
(see Section~\ref{sec:sim}) that the power at $\theta_1$ is generally
close to, but slightly less than, $1-\widetilde{\alpha}$.  The values
$\eps,\widetilde{\eps}$ are the fractions of type I and II error
probabilities ``spent'' at the first two stages, and in theory any
values $0<\eps,\widetilde{\eps}<1$ may be used. In practice, we
recommend using $0.2\le \eps,\widetilde{\eps}\le 0.8$ and we have
found that the power and expected sample size of the above adaptive
test vary very little with changes in $\eps,\widetilde{\eps}$. In
particular, the three examples in Section~\ref{sec:sim} use
$\eps=\widetilde{\eps}=1/3$, $(\eps,\widetilde{\eps})=(1/2,3/4)$, and
$\eps=\widetilde{\eps}=1/2$.  The factor $\rho_m$ in (\ref{eq2}) is a
small inflation of $n(\widehat{\theta}_m)$ to adjust for the
uncertainty in $\widehat{\theta}_m$. Lorden~\citeyearpar{Lorden83} gives an
asymptotic upper bound for $\rho_m$ as a function of $\theta_0$,
$\theta_1$, $\alpha$, and $\widetilde{\alpha}$. We advocate simply
fixing $\rho_m$ to a small maximum inflation that the practitioner is
comfortable with, and have found that $\rho_m=.05$ or $.1$ works well
in practice, which we use in the examples in Section~\ref{sec:sim}. As with $M$, the choice of $m$ is often determined by practical considerations like funding and duration. To aid such considerations or in the absence of them, if the practitioner has bounds $\underline{\theta}<\theta_0$ and $\overline{\theta}>\theta_1$ in mind (e.g., $\overline{\theta}$ might be the largest realistic treatment effect likely to be seen), then $m$ could be chosen to be $n(\underline{\theta})\wedge n(\overline{\theta})$, an approximation to Hoeffding's~\citeyearpar{Hoeffding60} lower bound for the smallest expected sample size of a test with error probabilities $\alpha, \widetilde{\alpha}$ at $\theta_0, \theta_1$ when $\theta=\underline{\theta}$ or $\overline{\theta}$.

The probabilities in (\ref{eq7})-(\ref{eq9}) can be computed by Monte
Carlo or recursive numerical integration, using normal approximations
to signed-root likelihood ratio statistics. Further details are given
in Section~\ref{subsec:numerical}.  The original idea to use (\ref{eq2}) as the
second-stage sample size and to allow the possibility of a third stage
to account for uncertainty in the estimate $\widehat{\theta}_m$ (and
hence $n_2$) is due to Lorden~\citeyearpar{Lorden83}, although his test uses
very conservative upper bounds on the error probabilities. Here we
have modified Lorden's test to control the type I error $\alpha$
exactly, and provided algorithms to implement the modified test.  It
can be shown that our three-stage test is asymptotically optimal: If
$N$ is the sample size of our three-stage test above, then
\begin{equation}
E_{\theta}(N)\sim
m\vee\left\{M\wedge\frac{|\log\alpha|}{I(\theta,\theta_0)\vee
  I(\theta,\theta_1)}\right\} \label{eq10} 
\end{equation}
as $\alpha+\widetilde{\alpha}\rightarrow 0$,
$\log\alpha\sim\log\widetilde{\alpha}$, $\rho_m\rightarrow 0$ and
$\rho_m\sqrt{m/\log m}\rightarrow\infty$; and if $T$ is the sample
size of any test of $H_0:\theta\le\theta_0$ whose error probabilities
at $\theta_0$ and $\theta_1$ do not exceed $\alpha$ and
$\widetilde{\alpha}$, respectively, then
\begin{equation}
E_{\theta}(T)\ge
(1+o(1))E_{\theta}(N) \label{eq25} 
\end{equation} simultaneously for all $\theta$. The proof uses
Hoeffding's~\citeyearpar{Hoeffding60} lower bound for $E_{\theta}(T)$ as in
\citeyearpar{Lorden83} and can be found in \citeyearpar{Bartroff07b}.

Since $\log\alpha\sim\log(\eps\alpha)$ as $\alpha\To 0$ for any fixed
$0<\eps<1$, the asymptotic formula for $E_\theta (N)$ in (\ref{eq10}) is
unchanged if one replaces the type I error probability $\alpha$ by a
fraction of it, and this is why Lorden~\citeyearpar{Lorden83} can use crude bounds of
the type above for the type I error probability. For values of the type I error
probability $\alpha$ (e.g., .05 or .01) commonly used in practice,
replacing $\alpha$ by $\alpha/10$, say, can substantially increase
$E_\theta(N)$. Note that our adaptive test keeps the error probability
at $\theta_0$ to be $\alpha$ (instead of less than $\alpha$) by using
Monte Carlo or recursive numerical integration to evaluate it, discussed in the next section.

\subsection{The Normal Case and Recursive Numerical Integration}
\label{subsec:numerical}

The thresholds $b$, $\widetilde b$, and $c$ can be computed by solving
in succession (\ref{eq7}), (\ref{eq8}), and (\ref{eq9}). Univariate
grid search or Brent's method~\citeyearpar{Press92} can be used to solve each
equation. Suppose the $X_i$ are $N(\theta, 1)$. Without loss of
generality, we shall assume that $\theta_0 = 0$. Since $I(\theta,
\lambda) = (\theta - \lambda)^2 / 2$, we can rewrite (\ref{eq7}) as
 \begin{equation}\begin{split}&\mbox{pr}_{\theta_1} \{S_m - m \theta_1 \leq - (2 \widetilde b
 m)^{1/2}\}\\
&+\mbox{pr}_{\theta_1} \{S_m - m \theta_1 > - (2 \widetilde b m)^{1/2},
 S_{n_2} - n_2 \theta_1 \leq - (2 \widetilde b n_2)^{1/2}\}
 =\widetilde\varepsilon  \widetilde \alpha\label{eq19}\end{split}\end{equation}
and (\ref{eq8}) and (\ref{eq9}) as
 \begin{align}&\mbox{pr}_0 \{S_m / \sqrt{2m} \geq b^{1/2}\}+ \mbox{pr}_0 \{\widetilde b^{1/2} < S_m / \sqrt{2m} < b^{1/2},
 S_{n_2} / \sqrt{2 n_2} \geq b^{1/2}\} = \varepsilon\alpha,\label{eq20}\\
 &\mbox{pr}_0\{\widetilde b^{1/2} < S_m / \sqrt{2m} < b^{1/2},
 \widetilde b^{1/2} < S_{n_2} / \sqrt{2 n_2} < b^{1/2}, S_M / \sqrt M
 \geq c^{1/2} \} = (1 - \varepsilon) \alpha.\label{eq21}\end{align} 
The probabilities involving $n_2$ can be computed by conditioning
on the value of $S_m/m$, which completely determines the value of
$n_2$, denoted by $k(x)$. For example, the probabilities under
$\theta=0$ can be computed via  
\begin{eqnarray}
&&\mbox{pr}_0\{S_{n_2} \geq (2 b n_2)^{1/2} | S_m = m x\} = \Phi\left(\frac{mx-[2bk(x)]^{1/2}}{[k(x)-m]^{1/2}}\right),\label{eq22}\\
&&\mbox{pr}_0 \{S_{n_2} \in dy, S_M \in dz | S_m = mx\} = \varphi_{k(x) -
  m} (y - mx) \varphi_{M - k(x)} (z-y) dydz, \label{eq11}
\end{eqnarray} where $\Phi$ is the standard normal c.d.f.\ and $\varphi_v$ is the $N(0, v)$ density function, i.e.,
$\varphi_v(w) = (2 \pi v)^{-{1/2}} \exp(-w^2 / 2v)$. The
probabilities under $\theta_1$ can be computed similarly. Hence
standard recursive numerical integration algorithms can be used to
compute the probabilities in (\ref{eq7})-(\ref{eq9}).

As an example, we compute the thresholds $b,\widetilde{b}$, and $c$
for the following adaptive test whose performance is studied in
Section~\ref{subsec:varknown}.  Here $M=120$, $\alpha=.025$, and we
want the power to be close to $1-\widetilde{\alpha}=.9$ at
$\theta=\theta_1=.3$. Setting $\eps=\widetilde{\eps}=1/3$ and
$\rho_m=.1$, we first find $\widetilde{b}$ by solving (\ref{eq19}),
which can be written as
$$\Phi(-[2\widetilde{b}]^{1/2})
+\int_{[2\widetilde{b}/m]^{1/2}}^\infty
\Phi\left(\frac{-m(x-\theta_1)-[2\widetilde{b}k(x)]^{1/2}}{[k(x)-m]^{1/2}}\right)\varphi_m(mx)mdx=\widetilde{\eps}\widetilde{\alpha}=\frac{0.1}{3}$$
by the analog of (\ref{eq22}) for $\theta=\theta_1$, where $\varphi_v$
is as in (\ref{eq11}). The integral is computed by numerical
integration and a few iterations of the bisection method gives
$\widetilde{b}=1.99$. This value is next used to find $b$ similarly by
solving (\ref{eq20}), which can be written as
$$\Phi(-[2b]^{1/2})+ \int_{[2\widetilde{b}/m]^{1/2}}^{[2b/m]^{1/2}}
\Phi\left(\frac{mx-[2bk(x)]^{1/2}}{[k(x)-m]^{1/2}}\right)\varphi_m(mx)mdx=\eps\alpha
=\frac{0.025}{3}$$ by (\ref{eq22}). The bisection method gives $b=3.26$, which
we in turn use to find $c$ by solving (\ref{eq21}), which is
$$\int_{[2\widetilde{b}/m]^{1/2}}^{[2b/m]^{1/2}}
\int_{[2\widetilde{b}k(x)]^{1/2}}^{[2bk(x)]^{1/2}}
\Phi\left(\frac{[cM]^{1/2}-y}{[M-k(x)]^{1/2}}\right)\varphi_{k(x)-m}(y-mx)
\varphi_m(mx)m dy dx=(1-\eps)\alpha=\frac{0.05}{3}$$ by (\ref{eq11}),
giving $c=2.05$.

\subsection{Multiparameter Extension}
\label{sec:multi}

Suppose $X_1, X_2,\ldots$ are independent $d$-dimensional random
vectors from a multiparameter exponential family $f_\theta (x) =
\exp\{\theta^T x - \psi(\theta)\}$ of densities.  The three-stage test
in Section~\ref{subsec:1param} can be readily extended to test $H_0 :
u(\theta) \leq u_0$, where $u$ is any smooth real-valued function.  As
in Section~\ref{subsec:1param}, $n_1 = m$ and $n_3 = M$. The stopping
rule of the three-stage test of $H_0 : u(\theta) \leq u_0$ is the same
as (\ref{eq4})-(\ref{eq6}) but with $n I(\widehat{\theta}_n,\theta_j)$
replaced by
\begin{equation}\label{eq34}\inf_{\theta:u(\theta)=u_j} n I(\widehat{\theta}_n,\theta),\end{equation} $j=0,1$,
where $u_1>u_0$ is the alternative implied by the maximum sample size $M$
and the desired type II error probability $\widetilde{\alpha}$; see \citeyearpar[Section~3.4]{Lai04}.  In
particular, the test stops and rejects $H_0$ at stage $i\le 2$ if
\begin{equation}\label{eq27}
n_i<M,\quad u (\widehat \theta_{n_i}) > u_0,\quad\mbox{and}\quad
\inf_{\theta: u(\theta)=u_0} n_i I(\widehat{\theta}_{n_i},\theta)\ge
b,\end{equation} which is analogous to (\ref{eq4}). Early stopping for futility (accepting $H_0$) can also occur at
stage $i \leq 2$ if
\begin{equation}\label{eq28}
n_i<M,\quad u(\widehat \theta_{n_i}) < u_1,\quad\mbox{and}\quad
\inf_{\theta: u(\theta)=u_1} n_i I(\widehat{\theta}_{n_i},\theta)\ge
\widetilde{b},\end{equation} which is analogous to (\ref{eq5}). The
test rejects $H_0$ at stage $i=2$ or $3$ if 
\begin{equation}\label{eq29}
n_i=M,\quad u(\widehat \theta_M) > u_0,\quad\mbox{and}\quad
\inf_{\theta: u(\theta)=u_0} M I(\widehat{\theta}_M,\theta)\ge
c,\end{equation} accepting $H_0$ otherwise.
The thresholds $b$, $\widetilde
b$, and $c$ are chosen to ensure certain type I and type II error probability
constraints that are similar to (\ref{eq7})-(\ref{eq9}) and are
computed by using the normal approximation to the signed-root
likelihood ratio statistic
$$\ell_n(\delta)=n\{\mbox{sign}(u(\widehat{\theta}_n)-\delta)\}
\{2\inf_{\theta: u(\theta)=\delta}
I(\widehat{\theta}_n,\theta)\}^{1/2}$$
under the hypothesis
$u(\theta)=\delta$; see \citeyearpar[p.~513]{Lai04}. Note that this normal
approximation can be used for the choice of $u_1$ implied by the
maximum sample size $M$ and the type II error probability
$\widetilde{\alpha}$. The sample size $n_2$ of the three-stage test is
given by (\ref{eq2}) with
\begin{equation}\label{eq30}
n(\theta) = \min\{|\log \alpha | / \inf_{\lambda : u (\lambda) = u_0}
I(\theta, \lambda), |\log \widetilde \alpha | / \inf_{\lambda : u(\lambda) =
  u_1} I(\theta, \lambda)\},\end{equation} which is a generalization
of (\ref{eq3}). Examples of the multiparameter case are given in
Section~\ref{subsec:varunknown} for normally distributed data with
unknown variance, and in Section~\ref{subsec:2bino} for two binomial populations.

\section{COMPARISON WITH OTHER TESTS}
\label{sec:sim}

\subsection{Normal Mean with Known Variance}
\label{subsec:varknown}

We consider the special case of normal $X_i$ with unknown mean
$\theta$ and known variance 1, and compare a variety of adaptive tests
of $H_0: \theta\le0$ in the literature with the tests proposed in
Section~\ref{subsec:1param}. In this normal setting,
$\widehat{\theta}_n=\overline{X}_n$ and
$I(\theta,\lambda)=(\theta-\lambda)^2/2$. It is widely recognized that
the performance of adaptive tests is difficult to evaluate and compare
because it depends heavily on the choice of first-stage and maximum
sample sizes, the number of groups (stages) allowed, and the parameter
values at which the tests are evaluated. For this reason, the tests
evaluated here use the same first-stage and maximum sample sizes,
except for a few illustrative examples discussed below. In addition,
we report a variety of operating characteristics for each test --
power, mean number of stages, and the 25th, 50th, and 75th percentiles
in addition to the mean of the sample size distribution -- over a wide
range of $\theta$ values. A comprehensive evaluation of adaptive and
group sequential tests like this has not appeared previously in the
literature.  We also include the uniformly most powerful FSS test with
the same maximum sample size and type I error probability $\alpha$,
which provides the appropriate benchmark for the power of any test of
$H_0$. Another relevant comparison -- especially given their
widespread use in clinical trials -- made here is with standard
(non-adaptive) group sequential tests having a similar number of
stages as the adaptive test.

To test $H_0: \theta \le 0$, Proschan and Hunsberger~\citeyearpar{Proschan95}
proposed a two-stage test, based on the conditional power criterion,
which uses the usual $z$-statistic but with a data-dependent critical
value to maintain the type I error at a prescribed level $\alpha$. The
test allows early stopping to accept (or reject) the null hypothesis
if the test statistic is below a user-specified upper normal quantile
$z_{p^*}$ (or above some level $k$) at the end of the first stage.
Choosing a data-dependent critical value is tantamount to multiplying
the $z$-statistic by a data-dependent factor and using a fixed
critical value. Li et al.~\citeyearpar{Li02} proposed to use the
$z$-statistic with a fixed critical value $c$, while still determining
the second-stage sample size by conditional power and maintaining the
type I error at $\alpha$. Their test stops after the first stage if
the test statistic falls below $h$ or above $k$. For each $h$ and
conditional power level, their test has a maximum allowable $k$, which
they denote by $k_1^*(h)$.  Fisher~\citeyearpar{Fisher98} proposed a
``variance spending'' method for weighting the observations so that
the type I error of his test does not exceed $\alpha$, despite its
data-dependent second-stage sample size that is given by the
conditional power criterion. To avoid a very large second-stage sample
size if the first-stage estimate of $\theta$ lies near the null
hypothesis, Shen and Fisher~\citeyearpar{Shen99} proposed early stopping due
to futility whenever the upper $100(1-\alpha_0)\%$ confidence bound
for $\theta$ falls below some specified alternative $\theta_1 > 0$

Table~1 compares these tests, a FSS test, and two standard group
sequential tests with the adaptive test described in
Section~\ref{subsec:1param}. The values of the user-specified
parameters of the tests are summarized in the list below.  The
user-specified parameters are chosen so that they have the same
first-stage sample size $m=40$ (except for the FSS test), maximum
sample size $M=120$ (except for SF$'$; see the last paragraph of this
section), type I error not exceeding $\alpha=.025$, and nominal
power (or conditional power level in the case of conditional power
tests) equal to .9.

\begin{itemize}
\item ADAPT: The adaptive test described in
  Section~\ref{subsec:1param} that uses $b=3.26$,
  $\widetilde{b}=1.99$, and $c=2.05$ corresponding to
  $\varepsilon=\widetilde{\varepsilon}=1/3$ in
  (\ref{eq7})-(\ref{eq9}), and $\rho_m=.1$ (see
  Section~\ref{subsec:numerical} for details).
\item FSS$_{120}$:  The FSS test having sample size 120.
\item OBF$_{PF}$, OBF$_{SC}$: O'Brien and Fleming's~\citeyearpar{OBrien79}
  one-sided group sequential tests having three groups of size 40.
  OBF$_{PF}$ uses power family futility stopping ($\Delta=1$ in
  \citeyearpar[Section~4.2]{Jennison00}) and OBF$_{SC}$ uses stochastic
  curtailment futility stopping ($\gamma=.9$ in
  \citeyearpar[Section~10.2]{Jennison00}). Both OBF$_{PF}$ and OBF$_{SC}$ use
  reference alternative $\theta_1=.3$; see below.
\item PH: Proschan and
Hunsberger's~\citeyearpar{Proschan95} test that uses $p^*=.0436$ and
$k=2.05$.
\item L: Li et 
al.'s~\citeyearpar{Li02} test that uses $h=1.63$ and
$k=k^*_1(h)=2.83$.
\item SF, SF$'$: Two 
versions of Shen and Fisher's~\citeyearpar{Shen99} test; SF uses
$\alpha_0=.425$ and SF$'$ uses $\alpha_0=.154$.
\end{itemize}

The tests are evaluated at the $\theta$ values where FSS$_{120}$ has
power .01, .025, .6, .8, .9, .95, and at $\theta=.15$, the midpoint of
$\theta=0$ and $\theta=\theta_1=.3$, the alternative implied by
$M=120$ since FSS$_{120}$
has power $1-\widetilde{\alpha}=.9$ there. This is also the
alternative used by the OBF tests for futility stopping. Each entry in
Table~1 is computed by Monte Carlo simulation with 100,000
replications. To compare tests $T$, $T'$ with type I error probability
$\alpha$ but with different type II error probabilities
$\widetilde{\alpha}_T(\theta)$, $\widetilde{\alpha}_{T'}(\theta)$
and expected sample sizes $E_\theta T$, $E_\theta T'$ at $\theta>0$,
Jennison and Turnbull~\citeyearpar{Jennison06a} defined the efficiency ratio
of $T$ to $T'$: \begin{equation}\label{eq26}R_\theta
  (T,T')=\frac{(z_\alpha+z_{\widetilde{\alpha}_T(\theta)})^2/E_\theta
    T}{(z_\alpha+z_{\widetilde{\alpha}_{T'}(\theta)})^2/E_\theta
    T'}\times 100,\end{equation} noting that
$(z_\alpha+z_{\widetilde{\alpha}_T(\theta)})^2/\theta^2$ is the sample
size of the FSS test with the same type I error probability and power
as $T$. Table~1 contains $R_\theta(T,N)$ for all tests $T$ and
$\theta>0$, where $N$ is the sample size of ADAPT.

\begin{center}
INSERT TABLE 1 ABOUT HERE
\end{center}

ADAPT has power comparable to FSS$_{120}$ at all values of $\theta$
while achieving substantial savings in sample size, as shown by the
percentiles and mean of the sample size. The three-stage OBF tests
have power comparable to ADAPT and FSS$_{120}$, but ADAPT has sample
size savings over the OBF tests, especially for larger $\theta>0$,
reflected by the efficiency ratio.  The mean number of stages (denoted
by \#) reveals that although ADAPT allows for the possibility of three
stages, most frequently it uses only one or two stages.

The conditional power tests PH, L, SF, and SF$'$ are underpowered at
values of $\theta>0$ in Table~1. In particular, PH, L, and SF all have
power less than .6 at $\theta_1=.3$, where ADAPT, FSS$_{120}$, and the
OBF tests have power around .9.  The lack of power of PH, L, and SF
shown by Table~1 is caused by stopping too early for futility.  For
example, the PH test stops for futility after the first stage if
$S_m/\sqrt{m}$ falls below $z_{p^*}=1.71$. But
$\mbox{pr}_{\theta_1}\{S_m/\sqrt{m}<1.71\}=.44$, well exceeding the
nominal type II error of .1. On the other hand, such stringent
futility stopping is necessary to control the sample size of
conditional power tests.  For example, the .025-level PH test that
stops for futility only when $\widehat{\theta}_{m}\le 0$ (i.e., with
$p^*=.5$) has expected sample size greater than $10^7$ at all values
of $\theta$ in Table~1, yet power less than .9 at $\theta_1$.  SF and
SF$'$ provide another example of this behavior.  Since these tests
stop for futility at the first stage when $S_m/m\le
\theta_1-z_{\alpha_0}/\sqrt{m}$, the choice of $\alpha_0$ determines
the maximum sample size.  For maximum sample size $M=120$, SF uses
$\alpha_0=.425$, a high rate of first-stage futility stopping which
results in small expected sample sizes, low power, and a reduced type
I error of .012, which is $\alpha=.025$ in the absence of futility
stopping. In contrast, SF$'$ uses less stringent futility stopping
with $\alpha_0=.154$ that corresponds to maximum sample size $5M=600$,
which results in a type I error closer to .025 and better power,
though it is still underpowered and its expected sample size exceeds
120 at $.2\le\theta\le .26$. The smallest $\alpha_0$ that does not
perturb the type I error of .025 of Shen and Fisher's test is
$\alpha_0=.039$, but the resultant test has expected sample size 1856
at $\theta=0$ and maximum sample size 52341.

\begin{center}
INSERT TABLE 2 ABOUT HERE
\end{center}

The efficiency ratios relative to ADAPT in Table~1 are all less than
100 with the exception of PH, L, and SF at $\theta=.15$, but it is not
clear that the efficiency ratio has much meaning in this case where
the power of these tests is so low.  For the other cases, it is
natural to ask if much more improvement is possible. A benchmark for
answering this question is provided by the optimal adaptive tests of
Jennison and Turnbull~\citeyearpar{Jennison06a,Jennison06b} that minimize the
expected sample size averaged over a collection of $\theta$ values,
subject to a given type I error probability and power level at a
prespecified alternative $\theta'$. Table~2 contains the expected
sample size of $T^*_k$, the $k$-stage test minimizing
\begin{equation}\label{eq33}[E_0(T)+E_{\theta'}(T)
  +E_{2\theta'}(T)]/3\end{equation} among all $k$-stage tests with
maximum sample size $M=120$, type I error probability $\alpha=.025$ and
power~.8 at $\theta'$, the alternative where FSS$_{100}$ has power~.8,
from~\citeyearpar[Table~III]{Jennison06b}. To this benchmark we compare
ADAPT with the same first group size $m=29$ as $T_3^*$, $M=120$,
$\theta_1$ fixed at $\theta'$, and $b=2.94$, $\widetilde{b}=.7$, and
$c=2.05$ corresponding to $\eps=1/2$, $\widetilde{\eps}=3/4$. Also
included in Table~2 is the optimal $k$-stage ``$\rho$-family'' group
sequential test (denoted by OGS($k$)) with $M=120$, groups
$2,\ldots,k$ of size $(M-m)/(k-1)$, and with $m$ and $\rho$ chosen to minimize
(\ref{eq33}). Jennison and Turnbull~\citeyearpar{Jennison06b} concluded that
OGS($k$) is a computationally easier alternative to $T_k^*$, and
Table~2 shows that their expected sample sizes are close at $\theta=0,
\theta',2\theta$.  Note that ADAPT has expected sample size close to
OGS(3) and $T^*_3$ even though the probability that ADAPT uses only 1
or 2 stages is 96.4\%, 83.1\%, and 98.4\% for $\theta=0,\theta'$, and
$2\theta'$, respectively, showing that ADAPT very often behaves like a
2-stage test. ADAPT has substantially smaller expected sample size
than $T_2^*$ and OGS(2), however. On the other hand, $T_4^*$ is more
efficient than ADAPT but this is due in part to its smaller first
group of $m=24$, afforded by its additional stage.  Here we have matched the first group
$m=29$ of ADAPT to the that of $T_3^*$ for the purpose of comparison, but
in practice there is flexibility in its choice of $m$. The $T_k^*$ and
OGS($k$) tests, on the other hand, are rigid in their choice of $m$
that is determined by dynamic programming from the prespecified alternative $\theta'$, about which
there may be some
uncertainty before the trial. 

Lokhnygina~\citeyearpar{Lokhnygina04}, who considers somewhat different
objective functions than (\ref{eq33}), 
has computed and plotted the data-dependent total sample size of the
optimal 2-stage design as a
function of the first stage sample mean $\overline{X}_m$. Her results
show the total sample size to be a unimodal function of
$\overline{X}_m$, peaking between 0 and $\theta'$. For comparison,
Figure~1 plots the function $n(\theta)$ (\ref{eq3}) in the sample size updating
rule (\ref{eq2}) of ADAPT for the setting of Table~2. A similar shape
is exhibited by Figure~2.2 of \citeyearpar{Lokhnygina04} on the total sample
size function (which is $m$ plus the second-stage sample size) of the
optimal two-stage test. This is not surprising because to be optimal,
the expected sample size cannot differ much from
Hoeffding's~\citeyearpar{Hoeffding60} lower bound, of which $n(\theta)$ is a
close approximation.  Figure~1 differs dramatically from the total
sample size function of any untruncated two-stage conditional
power rule which increases to
infinity as $\widehat{\theta}_m$ approaches $0$. Jennison and
Turnbull~\citeyearpar[p.~672]{Jennison06c} have also pointed this out and suggested
that this is a source of inefficiency of two-stage conditional power tests.

\begin{center}
INSERT FIGURE 1 ABOUT HERE
\end{center}

\subsection{Case of Unknown Variance}
\label{subsec:varunknown} The optimal
adaptive test $T_k^*$ and the optimal group sequential tests OGS($k$)
in Table~2 require the variance of the observations to be known. As
pointed out above, in practice there is often little information about
the sampling variability before the trial. Dynamic programming is
difficult to carry out for the optimal adaptive test when the
$X_1,X_2,\ldots$ are i.i.d.\ $N(\mu,\sigma^2)$ and both $\mu$ and
$\sigma$ are unknown, and no analog of $T_k^*$ has been developed in
this setting. However, the optimal group sequential tests OGS($k$) in
Table~2 can be modified for the present setting by applying their
error spending functions to the sequential $t$-statistics, as
described in \citeyearpar[Section~11.5]{Jennison00}, which are denoted by
OGS$^*(k)$. In this section we compare ADAPT with OGS$^*(k)$ and other
tests for a normal mean when the variance is unknown. In the
notation of Section~\ref{sec:multi}, $\theta=(\mu,\sigma)^T$,
$u(\theta)=\mu$, $u_0=0$, and the generalized likelihood ratio
statistic (\ref{eq34}) is
$$(n/2)\log\left[1+\left(\frac{\overline{X}_n-u_j}{\widehat{\sigma}_n}\right)^2\right],$$
where $\widehat{\sigma}_n$ is the MLE of $\sigma$. Denne and
Jennison~\citeyearpar{Denne00} proposed an adaptive group-sequential
extension of Stein's~\citeyearpar{Stein45} 2-stage $t$-test in which the
total sample size and stopping boundaries are updated at each stage as
a function of the current estimate of $\sigma^2$. Lai and
Shih~\citeyearpar{Lai04} introduced tests of composite hypotheses for a
multiparameter exponential family which use the same stopping rule
(\ref{eq27})-(\ref{eq29}) as ADAPT but with prespecified group sizes.
The expected number of stages, power and expected sample size of these
tests are given in Table~3 at various $(\mu,\sigma)$ values. All
tests use $m=34$ (with the exception of OGS$^*$(4)) and $M=120$ (with the
exception of DJ which has unbounded maximum sample size), the first
stage and maximum sample sizes of OGS(3) in
Section~\ref{subsec:varknown}, and nominal power levels $\alpha=.025$ and
$1-\widetilde{\alpha}=.8$ at $(\mu,\sigma)=(0,1)$ and $(\theta',1)$,
respectively, where $\theta'$ is as in
Section~\ref{subsec:varknown}. Other values of the user-specified
parameters of the tests are listed below.

\begin{itemize}
\item ADAPT: The adaptive test described in Section~\ref{sec:multi}
  with $u_1$ fixed at $\theta'$, $b=2.49$, $\widetilde{b}=.59$ and $c=2.7$ corresponding to
  $\eps=1/2$, $\widetilde{\eps}=3/4$.
\item OGS$^*(k)$: Jennison and
  Turnbull's~\citeyearpar[Section~11.5]{Jennison00} group sequential $t$-test
  with $k$ groups and the same $m$ and error spending function as that of
  OGS($k$) in Table~2: OGS$^*$(3) uses $\rho=.99$ and group sizes 34,
  43, 43; OGS$^*$(4) uses $\rho=1.13$ and group sizes 29, 30, 30, 31.
\item DJ: The adaptive 3-stage $t$-test of Denne and
  Jennison~\citeyearpar{Denne00} with $\rho=.99$, to match OGS$^*$(3). 
\item LS: Lai and Shih's~\citeyearpar[Section~3.4]{Lai04} group sequential
  test with group sizes 34, 43, 43, so that $m=34$ and $M=120$.
\end{itemize}

\begin{center}
  INSERT TABLE 3 ABOUT HERE
\end{center}

When $\sigma=1$, ADAPT, OGS$^*$, and DJ have similar power and
expected sample size properties, with ADAPT having the smallest
expected number of stages and smaller expected sample size than
OGS$^*$(3). LS has the highest power of the five tests, but highest
expected sample size too.  When $\sigma<1$ and $\mu=0$, ADAPT has
substantially smaller expected sample size than OGS$^*$(3) and even
OGS$^*$(4).  DJ has similar operating characteristics to ADAPT and LS when
$\sigma<1$. However, when $\sigma>1$, the expected sample size of DJ
becomes much larger than those of other tests because its total sample
size is chosen to be proportional to the estimate of $\sigma^2$ at the
end of the previous stage.  In all cases evaluated, ADAPT has the
smallest expected number of stages, less than 2 in each case, showing
that it most often behaves like a FSS or 2-stage test, as in the
variance known setting of Table~1.
 
\subsection{Coronary Intervention Study}
\label{subsec:2bino}

The National Heart, Lung and Blood Institute (NHLBI) Type II Coronary
Intervention Study~\citeyearpar{Brensike82} was designed to investigate the
cholesterol-lowering affects of cholesytyramine on patients with Type
II hyperlipoproteinemia and coronary artery disease.  Patients were
randomized into cholesytyramine and placebo groups, and coronary
angiography was performed before and after five years of treatment.
It was found that the disease had progressed in 20 of 57 in the
placebo group and 15 of 59 in the cholesytyramine group.  Proschan and
Hunsberger~\citeyearpar{Proschan95} and Li et al.~\citeyearpar{Li02} have considered
how this study could have been extended by using their two-stage tests
for the difference in two normal means with common unknown variance.
To apply these tests to the NHLBI study, they assumed the first-stage
sample size to be $58=(57+59)/2$ for the normal problem and used the
arcsine transformation so that the difference between the transformed
binomial frequencies, $p_1$ for the placebo group and $p_2$ for the
treatment group, is
approximately normally distributed; details are given in the next
paragraph. As an alternative we apply the three-stage test in
Section~\ref{sec:multi} to two binomial populations.  In the notation of
Section~\ref{sec:multi}, to test $H_0:p_2\le p_1$ we have
$\theta=(p_1,p_2)^T$, $u(\theta)=p_2-p_1$, $u_0=0$, and the test
statistic $\inf_{\theta: u(\theta)=\delta}
nI(\widehat{\theta}_n,\theta)$ in (\ref{eq27})-(\ref{eq29}) takes the
form
$$n\left\{\widehat{p}_{1,n}\log\left(\frac{\widehat{p}_{1,n}}{p_{\delta,n}}
  \right) +
  \widehat{q}_{1,n}\log\left(\frac{\widehat{q}_{1,n}}{1-p_{\delta,n}}\right)
  +\widehat{p}_{2,n}\log\left(\frac{\widehat{p}_{2,n}}{p_{\delta,n}+\delta}
  \right)
  +\widehat{q}_{2,n}\log\left(\frac{\widehat{q}_{2,n}}{1-p_{\delta,n}-\delta}
  \right)\right\},$$
where $\widehat{p}_{i,n}$ is the maximum
likelihood estimator of $p_i$ based on $n$ observations,
$\widehat{q}_{i,n}=1-\widehat{p}_{i,n}$, and $p_{\delta,n}$ is the
maximum likelihood estimator of $p_1$ under the assumption
$p_2-p_1=\delta$. The treatment and placebo groups are assumed to have
the same per-group sample size during interim analyses, following
Proschan and Hunsberger~\citeyearpar{Proschan95} and Li et al.~\citeyearpar{Li02}.

Letting $S_n$ denote the sum of independent normal random variables
with mean $\mu$ and variance 1, following a pilot study of size $m$
resulting in $S_m=s_m$, Proschan and Hunsberger's~\citeyearpar{Proschan95}
test chooses $n_2$ and critical value $c$ to satisfy the conditional
power criterion
\begin{equation}\label{eq31}\mbox{pr} \{S_{n_2}/n_2^{1/2}>c|S_m=s_m,
\mu=s_m/m^{1/2}\}\ge 1-\widetilde{\alpha}\end{equation} and
type I error constraint
\begin{equation}\label{eq32}\mbox{pr}_0\{S_{n_2}/n_2^{1/2}>c|S_m=s_m\}=\alpha.\end{equation} 
In order to solve for $n_2$ and $c$, a parametric form for the
probability in (\ref{eq32}) is assumed, which contains a
user-specified futility boundary $h$ and critical value $k$ for the
internal pilot. Li et al.~\citeyearpar{Li02} introduce a modification of
Proschan and Hunsberger's~\citeyearpar{Proschan95} test in which the critical
value $c$ is specified before the internal pilot study but $h$, $k$,
and $n_2$ are chosen to satisfy (\ref{eq31}) and (\ref{eq32}) after
the internal pilot study.  This modification allows approximations to
the probabilities in (\ref{eq31}) and (\ref{eq32}) to be used in lieu
of a specific parametric form.  For the coronary intervention study,
Proschan and Hunsberger~\citeyearpar{Proschan95} and Li et al.~\citeyearpar{Li02}
propose using these tests with the variance-stabilizing transformation
$S_n=(2n)^{1/2}\{\arcsin(\widehat{p}_{1,n}^{1/2})-\arcsin(\widehat{p}_{2,n}^{1/2})\}$.

\begin{center}
  INSERT TABLE 4 ABOUT HERE
\end{center}

Table~4 gives the power, per-group expected sample size, and
efficiency ratio (\ref{eq26}), using the normal approximation,
relative to ADAPT (for alternatives $p_2>p_1$) of the following tests
for various values of $p_1, p_2$ near $15/59=.254$ and $20/57=.351$,
the values observed in the NHLBI study~\citeyearpar{Brensike82}.

\begin{itemize}
\item L: Li et
al.'s~\citeyearpar{Li02} test with $h=1.036$, $k=1.82$, $c=1.7$,
$\alpha=.05$, conditional power level .8, and first-stage size
$m=58$.
\item PH: Proschan and
Hunsberger's~\citeyearpar{Proschan95} test with $h=1.036$, $k=1.82$,
$\alpha=.05$, conditional power level .8, and first-stage size
$m=58$.
\item ADAPT: The adaptive test described in a previous paragraph with
  $m=58$, $M=302$ (the maximum sample size of L), and thresholds
  $b=2.36$, $\widetilde{b}=1.1$, and $c=1.55$ corresponding to
  $\alpha=.05$, $\tilde{\alpha}=.2$, and $\eps=\widetilde{\eps}=1/2$.
\end{itemize}
All three
tests use the same first-stage size $m=58$. ADAPT matches the maximum
sample size $M=302$ of L, and the parameters of PH determine its
maximum sample size to be slightly larger at 354.
The actual power of L and PH is around 50\% for the values of $p_1$
and $p_2$ in Table~4 with $p_2-p_1=.1$, and is less than 50\% when
$p_1=.254$ and $p_2=.351$ where they were designed to have
conditional power 80\%. This is caused in part by premature stopping
for futility at the end of the first stage. Indeed, L and PH use the
same futility boundary and their probability of stopping at the end of
the first stage when $p_1=.254$ and $p_2=.351$ is .47, well exceeding
the nominal Type II error probability .2.  One might ask if a
conditional power test can avoid this phenomenon by using a larger
first-stage sample size so that the estimate
$\widehat{p}_2-\widehat{p}_1$ is near 0 less often after the first
stage when the true difference $p_2-p_1$ is substantially greater than
0. If the first-stage sample size of L is raised to 162 (raising the
maximum sample size to 1331), the resultant test has power 79\% when
$p_1=.254$ and $p_2=.351$, approximately equal the power of ADAPT.
However, the expected sample size of this version of L is 264 at this
alternative, compared to the expected sample size 213.1 of ADAPT.
Similar oversampling also occurs for the values of $p_1$ and $p_2$ in
Table~4 with $p_2-p_1>.1$, where the power of L and PH is closer to
the nominal conditional power level of 80\%, but the efficiency ratio
drops to around 75\%.

\section{DISCUSSION}
\label{sec:disc}

Most previous works in the literature on adaptive design of clinical trials
and mid-course sample size adjustments have focused on
two-stage designs whose second-stage sample size is determined by the
results from the first stage using conditional power. Although this
approach is intuitively appealing, it does not adjust for the
uncertainty in the first-stage parameter estimates that are used to
determine the second-stage sample size. This can result in substantial
power loss, as shown in Section~\ref{subsec:varknown}. Although Jennison and
Turnbull~\citeyearpar{Jennison03} and Tsiatis and Mehta~\citeyearpar{Tsiatis03} have
pointed out the inefficiency of this approach and advocate instead
using group sequential designs, their critique focuses on the use of
non-sufficient ``weighted'' test statistics and variability in the
interim estimate. Through our extensive
simulation studies we have shown that another problem with conditional
power methods in practice is potential lack of power, which results
from the difficulty in bridging conditional power with actual power
and in choosing a futility stopping rule.

In their recent survey of adaptive designs, Burman and
Sonesson~\citeyearpar{Burman06} pointed out that previous criticisms of the
statistical principles and properties of these designs may be
unconvincing in some situations when flexibility and not having to
specify parameters that are unknown at the beginning of a trial (like
the relevant treatment effect or variance) are more imperative than
efficiency or being powerful, whereas most efficient group sequential
designs require the prespecification of the relevant alternative and
variance, as in the case of the optimal adaptive tests of Jennison and
Turnbull~\citeyearpar{Jennison06a,Jennison06b}.  Moreover, conditional power
tests are easy to implement while optimal adaptive tests require
substantial dynamic programming computations.  The adaptive tests of
Section~\ref{sec:1param} combine the attractive features of both the
conditional power and group sequential tests. Rather than achieving
exact optimality at a specified collection of alternatives through
dynamic programming, they achieve asymptotic optimality over the
entire range of alternatives, resulting in near-optimality in
practice; see Section~\ref{subsec:varknown}. These tests are based on
efficient generalized likelihood ratio statistics which have an
intuitively ``adaptive'' appeal via estimation of unknown parameters
by maximum likelihood, ease of implementation, and freedom from having
to specify the relevant alternative (through the implied alternative)
that conditional power tests enjoy. As shown in
Section~\ref{sec:multi}, these generalized likelihood ratio statistics
and the associated adaptive tests can be readily extended to
multiparameter settings with nuisance parameters and they enjoy
near-optimality in these more complicated and realistic settings as
well; see Sections~\ref{subsec:varunknown} and \ref{subsec:2bino}.

The possibility of adding a third stage to improve two-stage designs
dated back to Lorden~\citeyearpar{Lorden83}, who used upper bounds for the
type I error probability that are overly conservative for applications
to clinical trials, which need to maintain the type I error
probability of the test at a prescribed level because of regulatory
and publication requirements; see the references in
Section~\ref{sec:intro}.  We have modified Lorden's three-stage test
by combining its basic features to preserve its asymptotic optimality
with those of Lai and Shih~\citeyearpar{Lai04} for efficient group sequential
designs. The adaptive test in Section~\ref{sec:1param} makes use of
the maximum sample size $M$ to come up with an implied alternative
which is used to choose the rejection and futility boundaries
appropriately so that the test does not lose much power in comparison
with the (most powerful) FSS test of the null hypothesis versus the
implied alternative. This idea has led to the superior power
properties of ADAPT in Table~1, comparable to those of the FSS test.
Moreover, the expected number of stages of ADAPT in Table~1 ranges
from 1.5 to 2.07 and is less than 2 for all cases in Table~3.
Therefore ADAPT is not much less convenient to run than the FSS test
(with only 1 stage), in contrast with group sequential tests with 3,
4, 5 or more interim analyses of the accumulated data. In practical
terms, this can provide substantial savings in the operational costs
of the trial by eliminating the need for data monitoring at interim
analyses since the updated sample size and stopping rule rule are
completely determined at the end of the pilot stage.

On the other hand, there are situations where adding an additional
stage or increasing the maximum sample size may be desired, as pointed
out by Cui, Huang, and Wang~\citeyearpar{Cui99} and Lehmacher and
Wassmer~\citeyearpar{Lehmacher99}. For example, \citeyearpar{Cui99} cites a study protocol,
which was reviewed by the Food and Drug Administration, involving a
Phase III group sequential trial for evaluating the efficacy of a new
drug to prevent myocardial infarction in patients undergoing coronary
artery bypass graft surgery.  During interim analysis, the observed
incidence for the drug achieved a reduction that was only half of the
target reduction assumed in the calculation of the maximum sample size
$M$, resulting in a proposal to increase the maximum sample size to
$N_{\max}$. The basic idea underlying the proposed test in
Section~\ref{sec:1param} can be easily modified to allow increase of
the maximum sample size from $M$ to no more than $N_{\max}$ after
the second stage, resulting in a test with at most four stages. The
type I error probability of the modified test can be computed
numerically by recursive integration or by Monte Carlo simulations, as
described in Section~\ref{sec:1param}.

\section*{ACKNOWLEDGMENTS}

Bartroff's work was supported by grant DMS-0403105 from the National Science
Foundation. Lai's work was supported by grants RO1-CA088890 and
DMS-0305749 from the National Institutes of Health and the National
Science Foundation.


\begin{thebibliography}{29}
\providecommand{\natexlab}[1]{#1}
\providecommand{\url}[1]{\texttt{#1}}
\providecommand{\urlprefix}{URL }

\bibitem[{Bartroff and Lai(2007)}]{Bartroff07b}
Bartroff, J. and Lai, T.~L. (2007).
\newblock Supplement to ``{E}fficient Adaptive Designs with Mid-Course Sample
  Size Adjustment in Clinical Trials''.
\newblock \url{http://www-rcf.usc.edu/~bartroff/research/adaptive_supp.pdf}.

\bibitem[{Brensike et~al.(1982)Brensike, Kelsey, Passamani, Fisher, Richardson,
  Loh, Stone, Aldrich, Battaglini, Moriarty, Marianthopoulos, Detre, Epstein,
  and Levi}]{Brensike82}
Brensike, J.~F., Kelsey, S.~F., Passamani, E.~R., Fisher, M.~R., Richardson,
  J.~M., Loh, I.~K., Stone, N.~J., Aldrich, R.~F., Battaglini, J.~W., Moriarty,
  D.~J., Marianthopoulos, M.~B., Detre, K.~M., Epstein, S.~E., and Levi, R.~I.
  (1982).
\newblock {NHLBI} Type {II} Coronary Intervention Study: {D}esign, Methods and
  Baseline Characteristics.
\newblock \emph{Controlled Clinical Trials} 3: 91--111.

\bibitem[{Burman and Sonesson(2006)}]{Burman06}
Burman, C.~F. and Sonesson, C. (2006).
\newblock Are Flexible Designs Sound? ({W}ith Discussion).
\newblock \emph{Biometrics} 62: 664--683.

\bibitem[{Chernoff(1972)}]{Chernoff72}
Chernoff, H. (1972).
\newblock \emph{Sequential Analysis and Optimal Design}.
\newblock Philadelphia: Society for Industrial and Applied Mathematics.

\bibitem[{Cui et~al.(1999)Cui, Hung, and Wang}]{Cui99}
Cui, L., Hung, H.~M., and Wang, S.~J. (1999).
\newblock Modification of Sample Size in Group Sequential Clinical Trials.
\newblock \emph{Biometrics} 55: 835--857.

\bibitem[{Denne and Jennison(2000)}]{Denne00}
Denne, J.~S. and Jennison, C. (2000).
\newblock A Group Sequential {t}-test with Updating of Sample Size.
\newblock \emph{Biometrika} 87: 125--134.

\bibitem[{Fisher(1998)}]{Fisher98}
Fisher, L. (1998).
\newblock Self-designing clinical trials.
\newblock \emph{Statistics in Medicine} 17: 1551--1562.

\bibitem[{Hoeffding(1960)}]{Hoeffding60}
Hoeffding, W. (1960).
\newblock Lower Bounds for the Expected Sample Size and the Average Risk of a
  Sequential Procedure.
\newblock \emph{The Annals of Mathematical Statistics} 31: 352--368.

\bibitem[{Jennison and Turnbull(2000)}]{Jennison00}
Jennison, C. and Turnbull, B.~W. (2000).
\newblock \emph{Group Sequential Methods with Applications to Clinical Trials}.
\newblock New York: Chapman \& Hall/CRC.

\bibitem[{Jennison and Turnbull(2003)}]{Jennison03}
Jennison, C. and Turnbull, B.~W. (2003).
\newblock Mid-Course Sample Size Modification in Clinical Trials Based on the
  Observed Treatment Effect.
\newblock \emph{Statistics in Medicine} 22: 971--993.

\bibitem[{Jennison and Turnbull(2006{\natexlab{a}})}]{Jennison06a}
Jennison, C. and Turnbull, B.~W. (2006{\natexlab{a}}).
\newblock Adaptive and Nonadaptive Group Sequential Tests.
\newblock \emph{Biometrika} 93: 1--21.

\bibitem[{Jennison and Turnbull(2006{\natexlab{b}})}]{Jennison06c}
Jennison, C. and Turnbull, B.~W. (2006{\natexlab{b}}).
\newblock Discussion on ``{A}re Flexible Designs Sound?''.
\newblock \emph{Biometrics} 62: 670--673.

\bibitem[{Jennison and Turnbull(2006{\natexlab{c}})}]{Jennison06b}
Jennison, C. and Turnbull, B.~W. (2006{\natexlab{c}}).
\newblock Efficient Group Sequential Designs When There are Several Effect
  Sizes Under Consideration.
\newblock \emph{Statistics in Medicine} 25: 917--932.

\bibitem[{Lai and Shih(2004)}]{Lai04}
Lai, T.~L. and Shih, M.~C. (2004).
\newblock Power, sample size and adaptation considerations in the design of
  group sequential clinical trials.
\newblock \emph{Biometrika} 91: 507--528.

\bibitem[{Lehmacher and Wassmer(1999)}]{Lehmacher99}
Lehmacher, W. and Wassmer, G. (1999).
\newblock Adaptive Sample Size Calculations in Group Sequential Trials.
\newblock \emph{Biometrics} 55: 1286--1290.

\bibitem[{Li et~al.(2002)Li, Shih, Xie, and Lu}]{Li02}
Li, G., Shih, W.~J., Xie, T., and Lu, J. (2002).
\newblock A Sample Size Adjustment Procedure for Clinical Trials.
\newblock \emph{Biostatistics} 3: 277--287.

\bibitem[{Lokhnygina(2004)}]{Lokhnygina04}
Lokhnygina, Y. (2004).
\newblock \emph{Topics in Design and Analysis of Clinical Trials}.
\newblock Ph.D. thesis, North Carolina State University.

\bibitem[{Lorden(1983)}]{Lorden83}
Lorden, G. (1983).
\newblock Asymptotic efficiency of three-stage hypothesis tests.
\newblock \emph{The Annals of Statistics} 11: 129--140.

\bibitem[{O'Brien and Fleming(1979)}]{OBrien79}
O'Brien, P.~C. and Fleming, T.~R. (1979).
\newblock A Multiple Testing Procedure for Clinical Trials.
\newblock \emph{Biometrics} 35: 549--556.

\bibitem[{Posch and Bauer(1999)}]{Posch99}
Posch, M. and Bauer, P. (1999).
\newblock Adaptive two stage designs and the conditional error function.
\newblock \emph{Biometrical Journal} 41: 689--696.

\bibitem[{Press et~al.(1992)Press, Flannery, Teukolsky, and
  Vitterling}]{Press92}
Press, N.~H., Flannery, B.~P., Teukolsky, S.~A., and Vitterling, W.~T. (1992).
\newblock \emph{Numerical Recipes in C: The Art of Scientific Computing}.
\newblock Cambridge University Press, 2nd edition.

\bibitem[{Proschan and Hunsberger(1995)}]{Proschan95}
Proschan, M. and Hunsberger, S. (1995).
\newblock Designed Extension Studies Based on Conditional Power.
\newblock \emph{Biometrics} 51: 1315--1324.

\bibitem[{Schmitz(1993)}]{Schmitz93}
Schmitz, N. (1993).
\newblock \emph{Optimal Sequentially Planned Decision Procedures}, volume~79 of
  \emph{Lecture Notes in Statistics}.
\newblock New York: Springer-Verlag.

\bibitem[{Shen and Fisher(1999)}]{Shen99}
Shen, Y. and Fisher, L. (1999).
\newblock Statistical Inference for Self-Designing Clinical Trials with a
  One-Sided Hypothesis.
\newblock \emph{Biometrics} 41: 190--197.

\bibitem[{Shih(2001)}]{Shih01}
Shih, W.~J. (2001).
\newblock Sample size re-estimation -- journey for a decade.
\newblock \emph{Statistics in Medicine} 20: 515--518.

\bibitem[{Stein(1945)}]{Stein45}
Stein, C. (1945).
\newblock A Two-Sample Test for a Linear Hypothesis Whose Power is Independent
  of the Variance.
\newblock \emph{The Annals of Mathematical Statistics} 16: 243--258.

\bibitem[{Tsiatis and Mehta(2003)}]{Tsiatis03}
Tsiatis, A.~A. and Mehta, C. (2003).
\newblock On the efficiency of the adaptive design for monitoring clinical
  trials.
\newblock \emph{Biometrika} 90: 367--378.

\bibitem[{Turnbull(2006)}]{Turnbull06}
Turnbull, B.~W. (2006).
\newblock Discussion on ``{S}tandard versus adaptive monitoring procedures: a
  commentary'' by {T}homas {R}.\ {F}leming.
\newblock \emph{Statistics in Medicine} 25: 3320--3325.

\bibitem[{Whitehead et~al.(2001)Whitehead, Whitehead, Todd, Bollard, and
  Sooriyarachi}]{Whitehead01}
Whitehead, J., Whitehead, A., Todd, S., Bollard, K., and Sooriyarachi, M.~R.
  (2001).
\newblock Mid-Trial Design Reviews for Sequential Clinical Trials.
\newblock \emph{Statistics is Medicine} 20: 165--176.

\end{thebibliography}

\def\cprime{$'$}

\end{document}